\documentclass[twocolumn,showpacs,preprintnumbers,amsmath,amssymb,superscriptaddress]{revtex4}

\usepackage{graphicx}
\usepackage{dcolumn}
\usepackage{bm}

\begin{document}

\preprint{APS/123-QED}

\title{Direct evidence for ferroelectric polar distortion in ultrathin lead titanate perovskite films}

\author{L. Despont}
\homepage{http://www.unine.ch/phys/spectro/homepage2.html}
\email{Laurent.Despont@unine.ch}
\affiliation{Institut de Physique, Universit\'{e} de Neuch\^{a}tel, CH-2000 Neuch\^{a}tel, Switzerland}%

\author{C. Lichtensteiger}
\affiliation{DPMC, Universit\'{e} de Gen\`{e}ve, 24 Quai Ernest-Ansermet, CH-1211 Gen\`{e}ve 4, Switzerland}%

\author{C. Koitzsch}
\affiliation{Institut de Physique, Universit\'{e} de Neuch\^{a}tel, CH-2000 Neuch\^{a}tel, Switzerland}%

\author{F. Clerc}
\affiliation{Institut de Physique, Universit\'{e} de Neuch\^{a}tel, CH-2000 Neuch\^{a}tel, Switzerland}%

\author{M. G. Garnier}
\affiliation{Institut de Physique, Universit\'{e} de Neuch\^{a}tel, CH-2000 Neuch\^{a}tel, Switzerland}%

\author{F. J. Garcia de Abajo}
\affiliation{Centro Mixto CSIC-UPV/EHU, 20080 San Sebasti\'{a}n, Spain}%

\author{E. Bousquet}
\affiliation{D\'{e}partement de Physique, Universit\'{e} de Li\`{e}ge, B-4000 Sart-Tilman, Belgium}%

\author{Ph. Ghosez}
\affiliation{D\'{e}partement de Physique, Universit\'{e} de Li\`{e}ge, B-4000 Sart-Tilman, Belgium}%

\author{ J.-M. Triscone}
\affiliation{DPMC, Universit\'{e} de Gen\`{e}ve, 24 Quai Ernest-Ansermet, CH-1211 Gen\`{e}ve 4, Switzerland}%

\author{P. Aebi}
\affiliation{Institut de Physique, Universit\'{e} de Neuch\^{a}tel, CH-2000 Neuch\^{a}tel, Switzerland}%

\date{\today}


\begin{abstract}
X-ray photoelectron diffraction is used to directly probe the
intra-cell polar atomic distortion and tetragonality associated with
ferroelectricity in ultrathin epitaxial PbTiO$_3$ films. Our
measurements, combined with \emph{ab-initio} calculations,
unambiguously demonstrate non-centro-symmetry in films a few unit
cells thick, imply that films as thin as 3 unit cells still preserve a
ferroelectric polar distortion, and also show that there is no thick paraelectric
dead layer at the surface.
\end{abstract}
\maketitle

\section{Introduction}
Theoretical developments and novel experiments in the area of
ferroelectrics have rapidly evolved over the last ten years,
allowing further progress in the understanding of this remarkable
phenomenon. In particular, ``nanoscale" ferroelectrics have
attracted considerable
attention~\cite{ahn04_1,lee05,was04,chu04,nau04}. The question of
the existence of a critical thickness, in other words whether or
not ferroelectricity can be maintained at reduced dimensions, is
amongst the most exciting topics of the field today, with very
active experimental~\cite{tyb99,str02,fon04,lic05} and theoretical
efforts~\cite{gho00,mey01,jun03_1}.

Probing ferroelectricity in thin films and nanostructures is a
difficult task, which requires advanced techniques. Among these,
scanning probe characterization based on piezoelectric microscopy
has allowed a ferroelectric ground state to be identified down to
40 \AA\ in thin Pb(Zr$_{0.2}$Ti$_{0.8}$)O$_3$ films~\cite{tyb99},
and x-ray studies on PbTiO$_3$ films suggested that 28 \AA\ films
are ferroelectric~\cite{lic05}. Dielectric and pyroelectric
response measurements have allowed ferroelectricity to be
identified in polymer films down to 10 \AA\ (two unit
cells)~\cite{bun98}. More recently, ultrahigh vacuum scanning
probe characterization based on electrostatic force microscopy was
used to study ferroelectricity in barium titanate nanowires with
diameters as small as 10 nm~\cite{yun02}. On insulating
substrates, lateral periodicity was observed via x-ray diffraction
in thin PbTiO$_3$ films down to 12 \AA\ and attributed to
alternately polarized domains~\cite{str02,fon04}. In all these
studies, however, properties {\it averaged} over the complete
ferroelectric structure were measured and no {\it local}
information on the atomic displacements was obtained.

In contrast, the photoemission based photoelectron diffraction
(XPD) used in our study presents two interesting characteristics:
it is naturally surface sensitive, due to an electron escape depth
of approximately 20 \AA\ (at the energy used here) ; and atomic
displacements within the unit cell can be directly probed,
allowing the non-centro-symmetric and tetragonal nature of
the crystal lattice to be directly demonstrated. This turns
out to be crucial for studying ultrathin films, and for
discriminating the behavior of the surface from that of the body of
the film.

The paper is organized as follows. In Section II, we characterize the ferroelectric distortion of PbTiO$_3$ and contrast it to the natural atomic relaxation appearing at surface and interface. In Sec. III A, we discuss aspects of the measurement methods  while, in Sec. III B we show the non-centro-symmetry of a 20 \AA\ thick film using XPD. The Sec. III C is dedicated to the tetragonality measurements as a function of film thickness with XPD using Pb as emitter, down to 4 \AA\ (one unit cell). We summarize and conclude in Sec. IV.

\section{Ferroelectric distortion versus surface relaxation}

\begin{figure}
\begin{center}
\includegraphics[width=8cm]{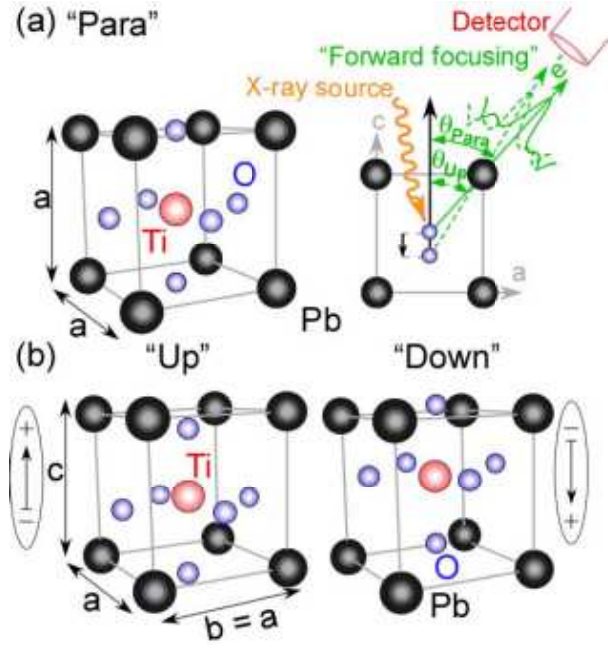}
\caption{\label{despontfig1} (a) Schematic of the
measurement setup. Core electrons have well-defined binding
energies, and their photoemission spectra exhibit characteristic
emission lines. By selecting a particular emission line,
photoelectrons from a given emitter can be chosen, thus probing
the local real-space environment of the emitter. The
interpretation is facilitated by forward focusing of electron flux
along the emitter-scatterer direction. The difference between two
``forward focusing" peak positions is directly related to a
modified emitter-scatterer direction, illustrated here by the
``para"- toward ``up"-state- O shift. Above 490$^\circ$C, bulk
PbTiO$_3$ has a cubic unit cell, with the O and Ti atoms centered
in the Pb cage (``para"-state). (b) Below 490$^\circ$C, bulk
PbTiO$_3$ has a tetragonal unit cell: two equivalent ferroelectric
configurations corresponding to two opposite polarizations,
``up"-state and ``down"-state. The displacements (in fractional
units) from cubic phase sites are 0.111 and 0.037 for O and Ti,
respectively~\cite{nel85}.}
\end{center}
\end{figure}

The studies were carried out on c-axis oriented perovskite
PbTiO$_3$ ultrathin films epitaxially grown on conducting
Nb-SrTiO$_3$ substrates. Above 490$^\circ$C, bulk PbTiO$_3$ is a
paraelectric insulator and a simple cubic perovskite structure
with a lattice parameter of 3.96 \AA\ (``para"-state, see
Fig.~\ref{despontfig1}(a)). In this structure, the Ti and O atoms
are in perfectly centro-symmetric positions with respect to the
surrounding Pb cage. At lower temperature, the material becomes
tetragonal and ferroelectric with a- and c-axis parameters of 3.90
\AA\ and 4.17 \AA\/, respectively ~\cite{nel85,jon97}, as
illustrated in Fig.~\ref{despontfig1}(b). The ferroelectric phase is
characterized by a non-centro-symmetric structure where the O and
Ti atoms are unequally shifted with respect to Pb. In a unit cell
with the polar c-axis along the $z$-direction, O and Ti move
either upwards or downwards (with a larger O displacement)
resulting respectively in a ``down"- and ``up"-polarized state, as
drawn in Fig.~\ref{despontfig1}(b).

\begin{figure}
\begin{center}
\includegraphics[width=8cm]{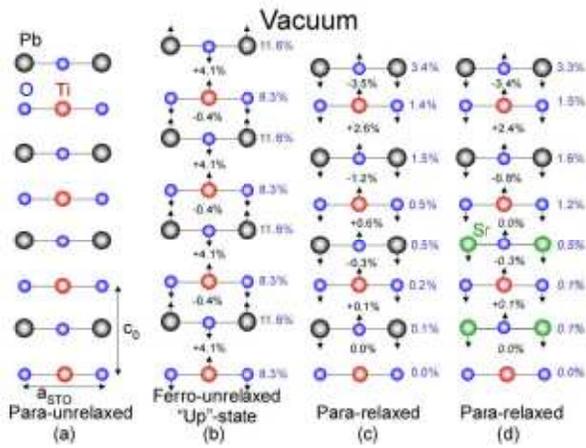}
\caption{\label{despontfig2} Schematic view of the atomic
displacements resulting from ferroelectric distortion and from
surface/interface relaxation in the uppermost layers of a
PbTiO$_3$ film epitaxially grown on SrTiO$_3$. In bulk PbTiO$_3$,
epitaxial strain produces a tetragonality at zero polarization
that can be estimated from the macroscopic elasticity theory as
$c_0/a_{STO}=1.03$. The ionic configuration resulting from the
truncation of such a strained bulk paraelectric state (a)
is considered as the reference structure. Freezing the bulk
ferroelectric distortion (as reported in Ref. \cite{nel85}) into
this reference structure results in the ``up"-state shown in
(b). Additionally, the natural ionic relaxation at the
surface in the paraelectric state has been computed from
first-principles (see text) both for a thick PbTiO$_{3}$ slab in
vacuum (c) and a SrTiO$_3$/(one unit cell) PbTiO$_3$/vacuum
stack (d). Numbers in black correspond to the change of
interlayer distances. Numbers in blue corresponds to the atomic
rumpling in each layer (cation-oxygen distance). All the values
are in \% of $c_0$, except those in italic that concern SrTiO$_3$
and are in \% of $a_{STO}$.}
\end{center}
\end{figure}

In the surface region (five top unit cells) that is probed by the
XPD technique, the evidence of such a polar atomic distortion could be
the signature of a ferroelectric ``up"- or ``down"-state but
may also arise from the natural atomic relaxation at the film
surface and interface already present in the paraelectric phase. A
proper interpretation of our data therefore requires independent
quantification of both effects. To that end, a {\it reference}
 configuration (``para-unrelaxed'') is defined in Fig.~\ref{despontfig2}(a):
it corresponds to the truncated bulk paraelectric structure of
PbTiO$_3$ with the in-plane lattice constant constrained to that
of SrTiO$_{3}$ ($a_{STO} = 3.90$ \AA) and a consequent
tetragonality $c_0/a_{STO} = 1.03$~\cite{lic05}.
Fig.~\ref{despontfig2}(b) (``ferro-unrelaxed'') shows the atomic
distortion of the ``up"-state as determined for bulk tetragonal
PbTiO$_3$ by Nelmes and Kuhs in Ref. \cite{nel85} with $c = 4.17$
\AA . In order to quantify the surface relaxation, density
functional theory calculations~\cite{abinit} were performed within
the local density approximation (LDA) using the
ABINIT~\cite{gon02} package. Two different supercells were
considered~: a thick PbTiO$_{3}$ slab in vacuum
(Fig.~\ref{despontfig2}(c)) and a SrTiO$_3$/(one unit cell)
PbTiO$_3$/vacuum stack~\cite{unitcell}(Fig.~\ref{despontfig2}(d)).
Insulating SrTiO$_3$ was considered in our simulations since Nb
doping is not presently affordable at the first-principles
level~\cite{STO}. To reproduce the substrate clamping effect, the
in-plane lattice constant was fixed to the relaxed a-axis value of
bulk SrTiO$_{3}$~\cite{acell}. The atomic positions were then
relaxed until the maximum residual atomic force was smaller than
40 meV/\AA\/. Our calculations were restricted to $(1 \times 1)$
surface periodicity and did not allow for an eventual
antiferrodistortive (AFD) $c(2 \times 2)$
reconstruction~\cite{bun05,sep05}. The latter is not excluded but,
as discussed later, was not evidenced on our films at room
temperature. The distortions in the upper half of each supercell
are reported in Fig.~\ref{despontfig2}(c), (d) (``para-relaxed"
state). For easy comparison with the experiment, because of the
typical LDA underestimate of the lattice constant~\cite{acell},
the values are given as a percentage of $c_0$. The magnitudes of
the ferroelectric and surface relaxation effects can now be
compared. First, the cation-oxygen displacements due to
ferroelectricity (11.6\% - 8.3\% of $c_0$,
Fig.~\ref{despontfig2}(b)) are significantly larger than the
displacements due to surface relaxation/rumpling (3.4\% - 1.4\% of
$c_0$, Fig.~\ref{despontfig2} (c) and 3.3\% - 1.5\% of $c_0$,
Fig.~\ref{despontfig2}(d)). Second, the mean layer displacement for
the ``up"-state (Fig.~\ref{despontfig2}(b)) and for the surface
relaxation (Figs.~\ref{despontfig2}, (c) and (d)) are opposite. Third,
the surface relaxation and rumpling effects are globally
unaffected by the film thickness (Figs.~\ref{despontfig2}, (c) and
(d)) and their amplitude decays very quickly in the interior of the
film : they are already negligible two unit cells below the
surface. This implies that the XPD measurements will be probing
both the narrow relaxed surface region and a few unit cells below,
essentially not affected by the surface relaxation.

\section{Experimental results and discussion}
\subsection{Experimental details }

The samples used in this study are epitaxial, c-axis oriented
PbTiO$_3$ thin films grown on conducting (001) Nb-SrTiO$_3$
substrates using off-axis radio-frequency magnetron
sputtering~\cite{eom90,lic04}. Topographic measurements using
atomic force microscopy (AFM) showed that these films are
essentially atomically smooth, with a root-mean-square roughness
between 2 and 6 \AA\ over a 10 $\times$10 $\mu$m$^2$ area. Room
temperature x-ray diffraction measurements, for films with
thickness $\geq $ 28 \AA\, allowed us to precisely determine the
thickness and the c-axis parameter of the films, and to confirm
their epitaxial ``cube-on-cube" growth.

After growth and characterization, the films were transferred
\emph{ex-situ} to a modified Vacuum Generators ESCALAB Mk II
photoelectron spectrometer. The XPD measurement system comprises a
hemispherical electron energy analyzer with a three-channel
detector, an x-ray photon source with two possible energies
($h\nu$ = 1253.6 eV and 1740 eV for MgK$\alpha$ and SiK$\alpha$
radiation, respectively), and a computer-controlled two-axis
goniometer capable of rotating the photoelectron emission angle
over the full hemisphere above the surface~\cite{ost91,nau93_1}.

The local geometry around a selected atom can be probed by performing an intensity versus emission-angle
 scan of a chosen photoemission line.
Because of the chemical sensitivity of photoemission, a given atom
type is then chosen by selecting one of its core levels. The
outgoing photoemitted electrons exhibit a strongly anisotropic
angular intensity distribution. This angular distribution is due
to the interference of the directly emitted photoelectron wave
with the scattered electron waves. The analysis of the
interference (or diffraction) patterns is facilitated by the
so-called ``forward focusing" effect taking place for photoelectron
kinetic energies greater than $\approx$ 0.5 keV. When considering
a row of atoms, scattering at the first few atoms along this row
focuses the electron flux in the emitter-scatterer direction (for
a review see Ref. \cite{ege90,fad90}). This enhancement of the
intensity in the emitter-scatterer direction is schematically
illustrated by the green curve in Fig.~\ref{despontfig1}(a) (right
part) for the centro-symmetric ``para"-state (continuous line) and
the ``up"-state (dotted line). The forward focusing effect is
further amplified for electron scattering by heavy atoms. In a
semi-classical picture this can be understood as the focusing of
the electron wave by the high number of protons in high atomic
number atoms~\cite{ege90}. Note that, despite the forward focusing
effect, the experimentally measured angles are sensitive to
multiple interferences, refraction and possible anisotropic atom
vibrations at the surface. In the present case of PbTiO$_3$, Pb
scattering is highly dominant compared to the scattering by other
elements~\cite{des05}. As a first step, in order to probe the
non-centro-symmetry, O was chosen as emitter-atom (O 1s core
level, E$_{kin}$ = 724.1 eV), since it has the largest
displacement~\cite{nel85} and has Pb scatterers as nearest
neighbors (see Fig.~\ref{despontfig3}). However, the O
contribution from the Nb-SrTiO$_{3}$ substrate becomes
non-negligible for films thinner than the photoelectron inelastic
mean free path, making the study of films thinner than 20 \AA\
more difficult. As a second step, in order to probe the
tetragonality of the films, i. e., the $c/a$ ratio of the Pb
lattice (related to the polarization via the polarization-strain
coupling as discussed below and in details in Ref.~\cite{lic05}),
Pb was chosen as emitter-atom (Pb 4f$_{7/2}$ core level, E$_{kin}$
= 1115.5 eV), and Pb-Pb forward focusing directions were used.
Since Pb atoms are absent from the substrate, this study can be
done down to a
 monolayer of ferroelectric material.

\subsection{Non-centro-symmetric position of oxygen atoms}

\begin{figure}
\begin{center}
\includegraphics[width=8cm]{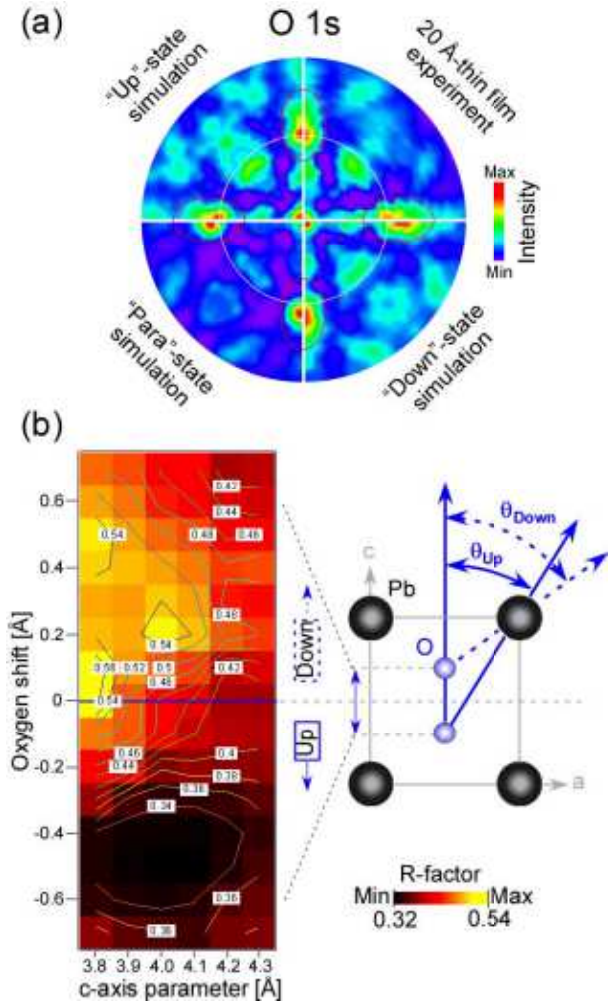}
\caption{\label{despontfig3} (a) Stereographic projection
of experimental and theoretical O 1s emission line intensities on
quarter hemispheres for four different cases: experimental data
for a 20 \AA\ thin film and theoretical simulations taking into
account ``up"-, ``para"- and ``down"-state structures. Normal
emission corresponds to the center of the plot and grazing
emission ($\theta = 70^\circ$) to the outer border. (b)
R-factor calculation to quantify the agreement between experiment
and simulations for different structures, where a low R-factor
corresponds to good agreement. A cut in the (100)-plane containing
Pb and O atoms is shown to facilitate the discussion. A downward
shift of Pb and O atoms optimizes the fit.}
\end{center}
\end{figure}

First, considering oxygen as the emitter atom, fully automated
computer code for calculating electron diffraction in atomic
clusters (EDAC) via multiple scattering~\cite{gar01}, based on the
muffin-tin potential approximation~\cite{pen74} was used to
calculate the XPD pattern. Fig.~\ref{despontfig3}(a) shows four O 1s
core level emission (E$_{kin}$ = 724.1 eV) interference patterns.
One is the measurement made on a 20 \AA\ thin film while the three
others are multiple scattering EDAC calculations of the ``up"-,
``down"- and ``para"-state. Intensities are plotted in a
stereographic projection with the center corresponding to normal
emission (polar angle $\theta = 0^\circ$) and the outer
 border corresponding to grazing emission ($\theta = 70^\circ$).
The strongest intensities (surrounded by red ellipses) correspond
to the scattering of O 1s photoelectrons by Pb nearest neighbours
(see Fig.~\ref{despontfig3}(b)). The white circle is a guide to the
eye indicating the polar angle of maximum intensity for the
measured interference pattern. It is evident that the polar angle
position of this peak, which is directly linked to
 the O-Pb directions, is perfectly reproduced by the ``up"-state calculation while the ``para"- and
the ``down"-state simulations predict a different position.
The ``up"-state (down-shifted O position) corresponds to a
smaller polar emission angle ($\theta_{up}$ in
Fig.~\ref{despontfig3}(b)) appearing closer to normal emission
(center of the interference pattern in Fig.~\ref{despontfig3}(a)).
Such measurements have also been performed on films with
thicknesses of $\sim$ 500, 200, 100, 60, 44 and 28 \AA\/, and all
perfectly reflect the characteristics of the
``up"-state \footnote{The presence of alternating 180$^{\circ}$
domains in our films can be ruled out from the XPD measurement
because the two characteristic ``up" and ``down"-state ``forward
focusing" peaks are not observed simultaneously in the
experimental diffractogram, Fig.~\ref{despontfig3}(a). This
contrasts with the results of Fong et al. \cite{fon04} on a
insulating substrates.}.

These conclusions, drawn from visual inspection of the
interference patterns locally around the intensity maximum
(Fig.~\ref{despontfig3}(a)), are confirmed by a global matching
approach using a reliability (R)-factor to evaluate the quality of
the fit between
 the complete experimental interference pattern data and theory (Fig.~\ref{despontfig3}(b)). The c-axis lattice
constant value and the O and Ti shifts are the adjustable
structural parameters. A cut in the (100) plane, containing Pb and
O atoms, is shown to facilitate the discussion. In the
calculation, O and Ti atoms
 are moved together and the dipole is continuously changed from the ``down"-state to the ``up"-state,
crossing over the ``para"-state. The best fit corresponds to the
minimal R-factor value, which is reached when O and Ti atoms are
shifted below the centro-symmetric position (parameters used for
the ``up"-state simulation in Fig.~\ref{despontfig3}(a)), with an
R-factor value of $\approx$ 0.34. In comparison, for the same
c-axis parameter but the opposite O and Ti atom shifts (parameters
used for the ``down"-state simulation in Fig.~\ref{despontfig3}(a)),
the calculation gives a much higher R-factor of $\approx  0.47$.
In between (zero O shift), in the centro-symmetric ``para"-state,
the R-factor is $\approx 0.45$ (parameters used for the
``para"-state simulation in Fig.~\ref{despontfig3}(a)).

It is important to note that surface relaxation and rumpling,
neglected here, cannot weaken our conclusions; indeed they would
give a picture resembling a ``down"-state, the corresponding O-Pb
atoms being shifted in the opposite direction than what is
observed (see Fig.~\ref{despontfig2}(c)). Also, the possibility of a
surface AFD reconstruction was explored without finding evidence
for it in our room temperature experiments.

This R-factor analysis therefore quantitatively confirms the
observations made in Fig.~\ref{despontfig3}(a), namely that the
measured interference pattern is best simulated with the ``up"-
state. This demonstrates unambiguously that, for a film as thin as
20 \AA\,  the O atoms have a non-centro-symmetric position in the
Pb cage corresponding to a non-vanishing spontaneous polarization.

Let us emphasize that piezoelectric AFM measurements performed
after the XPD experiments on the thickest films ($\sim$ 500 \AA )
confirmed the uniform``up"-state polarization while a uniform
``down"-state polarization had been initially found for the same
films just after growth.  This confirms the {\it monodomain}
character of the as grown sample and also indicates that the films
are uniformly switched from ``down" to ``up"-state by exposure to
our conventional x-ray source, attesting for the switchable
character of the polarization. The details behind the switching
are presently not known, but we believe that it occurs at the
initial stage of the experiment while the measurement itself is
essentially done in zero field. In fact, our results do not depend
on the x-ray intensity, proving that the films are in equilibrium
state during the measurements \footnote{The influence of the x-ray
intensity on the tetragonality was measured by investigating 
different x-ray powers. A modification of the
tetragonality would have indicated, via the polarization-strain
coupling, a variation of the spontaneous polarization. However no
such modification was found.}. As discussed below, the
agreement between the tetragonality deduced from x-ray diffraction
and XPD also suggests that the measurements are performed in
similar conditions.

\subsection{Tetragonality via lead emission}

\begin{figure}
\begin{center}
\includegraphics[width=8cm]{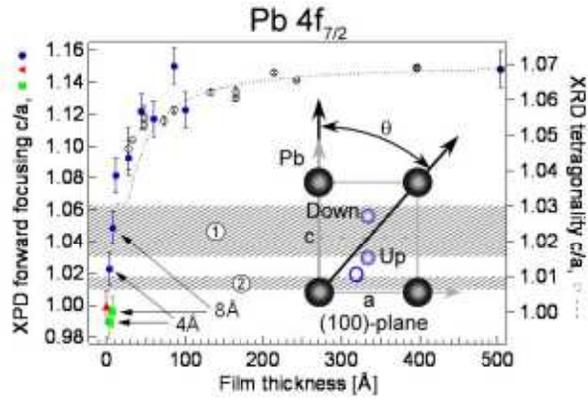}
\caption{\label{despontfig4} Tetragonality ($c/a$ ratio) as a
function of film thickness for PbTiO$_3$ films (blue circles), for
the Nb-SrTiO$_3$ substrate through PbTiO$_3$ films
 of one and two unit cells (green squares) and the Nb-SrTiO$_3$ substrate surface (red triangle). Black
open circles are x-ray diffraction data and the black dotted curve
is the result from model Hamiltonian calculations from
Ref. \cite{lic05}. Ribbons, labeled 1 and 2, indicate regions as discussed in the text. The left scale indicates the c/a values
extracted from the XPD experiment forward focusing directions
using $\theta$. The inset shows the
crystallographic plane used to extract the $c/a$ ratio. Pb-Pb
scattering dominates compared to Pb-O scattering allowing $\theta$
to be determined via Pb-Pb forward focusing.}
\end{center}
\end{figure}

In a second step, considering Pb as the emitter atom, XPD was used
to determine the tetragonality. As demonstrated in
Ref.~\cite{lic05}, below 200 \AA\ the tetragonality decreases as
the film thickness decreases. This decrease is a consequence of
the strong polarization-strain coupling in PbTiO$_3$ and a
signature of a reduced polarization in thin films. In
Ref.~\cite{lic05}, this polarization reduction was attributed to
imperfect screening of the depolarizing field~\cite{jun03_1}. With
XPD, using Pb as emitter, the tetragonality was measured down to
the unit cell level as shown in Fig.~\ref{despontfig4}. The
absolute values of  $c/a$, deduced from the forward focusing
angles, are particulary large. This might reflect
 a strong enhancement of the polarization in the
probed surface region (of the order of $80\% $ for $c/a = 1.15$,
from the PTO polarization-strain coupling), even larger than in
the theoretical prediction of Ref.~\cite{gho00}. However, as
previously stated, we are not necessarily measuring the precise
atom-atom directions and the anomalously large forward focusing $c/a$ might
also be related to other effects (anisotropic atom
vibrations at the surface, refraction and multiple scattering
interferences). Therefore a comparison
 with x-ray diffraction (XRD)~\cite{lic05} must be done at the relative level (Fig.~\ref{despontfig4},
left and right scale).

To study the evolution of the tetragonality as a function
of the film thickness, the measured XPD values are compared to the $c/a$ values obtained by XRD.
 The XPD measurement on Fig.~\ref{despontfig4}
confirms the evolution of c/a obtained from x-ray measurements in
Ref.~\cite{lic05} and agrees with the theoretical prediction
(dashed curve) relying on the suppression of polarization due to
imperfect screening of the depolarizing field~\cite{lic05}. The
similar thickness dependence for the XPD (very surface sensitive)
and the x-ray measurements (average on the whole film) implies
that the polarization evolves at the surface in the same way as at
the interior of the film and that there is no thick paraelectric
dead layer at the surface.  In addition, the XPD tetragonality
measurement shows a continuous decrease of tetragonality down to
the thickness of one unit cell~\cite{unitcell}. Two ribbons are
drawn  in Fig.~\ref{despontfig4}, labeled with 1 and 2. They
indicate the regions within which $c/a$ values of 1.03 and 1.01
are crossed with respect to both c/a scales. For film thicknesses
above two unit cells, the $c/a$ values are larger than 1.03, the
value expected at the bulk level for the paraelectric phase
(resulting from the mechanical constraint imposed by the
substrate, see also Fig.~\ref{despontfig2}(a)). This observation
directly implies, via the polarization-strain coupling, that the
films still have a finite -although progressively reduced-
spontaneous polarization. At thicknesses of one or two unit
cells, as can be seen on Fig.~\ref{despontfig4}, c/a drops even
more, reaching a value close to 1.01 for the one unit cell thick
film~\cite{unitcell}. This further decrease highlights that
macroscopic elasticity no longer applies at such thicknesses where
the interlayer atomic distances are affected by surface relaxation
and rumpling as shown by the \emph{ab-initio} calculations
(Fig.~\ref{despontfig2}(d)). The measured tetragonality agrees with
the computed value of 1.01 for the one unit-cell thick relaxed
paraelectric film suggesting the absence of any additional
ferroelectric distortion at this thickness.

\section{Conclusion}

This study thus directly demonstrates non-centro-symmetry,
unambiguously a result of ferroelectricity in PbTiO$_3$ thin films
down to 20 \AA . The measurements of the tetragonality, with a
continuous decrease down to the bare substrate, show that even
extremely thin films (3 unit cells) have a $c/a$ value larger than
1.03, attesting for the presence of a non-vanishing spontaneous
polarization at this thickness scale. As the film thickness is
reduced to a single unit cell, the experiments, together with
calculations, strongly suggest that both non-centro-symmetry and
tetragonality are governed by surface effects, giving rise for our
geometry to a polar relaxed structure but without switchable
ferroelectric distortion.

\section*{Acknowledgements}
We would like to thank M. A. Van Hove and C. Battaglia for helpful discussions, P. Paruch for careful reading of the manuscript, and the whole Neuch\^atel workshop and electric engineering team for efficient technical support. This project has been supported by the Swiss National Science Foundation through the National Center of Competence in
Research ``Materials with Novel Electronic Properties-MaNEP", the European Network of Excellence FAME and the VolkswagenStiftung.


\begin{thebibliography}{10}
\bibitem{abinit}
 Teter norm conserving pseudopotentials were used with Pb (5d, 6s, 6p), Sr
  (4s, 4p, 5s), Ti (3s, 3p, 3d, 4s) and O (2s, 2p) treated as valence states, a
  plane-wave energy cutoff of 35 Ha and 6x6x1 k-point sampling of the Brillouin
  zone.
\bibitem{unitcell}
 Note that a film with thickness of one unit cell corresponds to two Pb-O and
  one Ti-O layers.
\bibitem{STO}
 The Nb doping should have a minor effect on the relaxation at the
  paraelectric level except for ultrathin films for which the metallic
  character of the surface could allow a larger polar relaxation of the
  PbTiO$_3$ surface.
\end{thebibliography}

\end{document}